\documentstyle[aps,12pt,floats,epsf,tighten]{revtex}
\newcommand{\met}{$\rlap{\kern0.25em/}E_T$}
\newcommand{\gmet}{$\gamma\rlap{\kern0.25em/}E_T$}
\newcommand{\ggmet}{$\gamma\gamma\rlap{\kern0.25em/}E_T$}
\newcommand{\gmetjj}{$\gamma\rlap{\kern0.25em/}E_T+\ge 2\ {\rm jets}$}
\newcommand{\gmetnj}{$\gamma\rlap{\kern0.25em/}E_T+n\ {\rm jets}$}
\newcommand{\gjj}{$\gamma+\ge 2\ {\rm jets}$}
\newcommand{\emetjj}{$e\rlap{\kern0.25em/}E_T+\ge 2\ {\rm jets}$}
\begin{document}
\rightline{Fermilab-Conf-98/174-E}
\vspace*{2.0cm}

\begin{center}
  {\Large\bf New New-Phenomena Results from D\O}~\footnote{Presented at the 1998
  Rencontres de Physique de la Vall\'ee d'Aoste on Results and Perspectives in
  Particle Physics, La Thuile, Italy, March 1--7, 1998.}

  \vspace*{2.0cm}
  Jianming Qian \\
  \vspace*{0.5cm}
  {\sl Department of Physics, Univ. of Michigan, Ann Arbor, MI 48109, USA}\\
  \vspace{0.2cm}
  {\sl E-mail: qianj@umich.edu} \\
  \vspace{.5cm}
  {\sl (for the D\O\ Collaboration)}

  \vfil
  \begin{abstract}
    We have searched for diphoton events (\ggmet) with large missing transverse 
    momentum, $\gamma\rlap{\kern0.25em/}E_T$ events (\gmetjj) with two or more 
    jets, and diphoton events ($\gamma\gamma$) with high transverse energies in 
    $p\bar p$ collisions at $\sqrt{s} = 1.8$~TeV using approximately 
    100~pb$^{-1}$ of data collected with the D\O\ detector at the Fermilab 
    Tevatron in 1992--1996. No excess of events beyond the expected backgrounds 
    is observed. The null results are interpreted in supersymmetric models with
    a dominant $\tilde\chi^0_2\rightarrow\gamma\tilde\chi^0_1$ decay and in 
    terms of Dirac pointlike monopole production.
  \end{abstract}
\end{center}

\newpage
\section{Introduction}
The D\O\ detector collected a data sample corresponding to an integrated 
luminosity of approximately 100~pb$^{-1}$ during the 1992--1996 Tevatron Run at 
$\sqrt{s}=1.8$~GeV. The detector consists of three major components: 
a non-magnetic tracking system including a transition radiation detector 
in the central region, a liquid-argon calorimeter, and a toroidal magnetic 
muon spectrometer. A detailed description of the D\O\ detector can be found 
in Ref.~\cite{dzero}. The Tevatron $p\bar{p}$ collider is ideal for 
studying high $p_T$ phenomena. It is a natural place to search for new 
physics beyond the Standard Model. In this talk, the results of the searches 
for supersymmetry and Dirac monopoles are presented.

\section{Search for Supersymmetry Using Photons}
Supersymmetry~\cite{susy} is a generalization of space-time symmetry. 
It introduces for every Standard Model particle a supersymmetric partner 
that differs in spin by $1/2$. ${\cal R}$-parity conservation requires that 
supersymmetric particles be produced in pairs and that the lightest 
supersymmetric particle~(LSP) be stable.

The minimal supersymmetric standard model~(MSSM) is the simplest 
supersymmetric model. In the MSSM, the Gaugino-Higgsino sector
(excluding gluinos) is described by four parameters: $M_1$, $M_2$,
$\mu$, and $\tan\beta$, where $M_1$ and $M_2$ are the $U(1)$ and 
$SU(2)$ gaugino mass parameters, $\mu$ is the Higgsino mass parameter,
and $\tan\beta$ is the ratio of the vacuum expectation values of the 
two Higgs doublets. The Gaugino-Higgsino mixing gives in 
four neutral mass eigenstates (neutralinos $\tilde\chi^0_i,\ i=1,2,3,4$) 
and two charged mass eigenstates (charginos $\tilde\chi^\pm_j,\ j=1,2$) 
whose masses and couplings are fixed by the above four parameters.
Supersymmetric models~\cite{gmsb,kane} predicting photon production have 
been proposed as possible explanations of a recent event~\cite{cdf} reported 
by the CDF collaboration. Within the MSSM, it has been shown that the radiative 
decay of $\tilde\chi^0_2\rightarrow\tilde\gamma\chi^0_1$ dominates in 
some regions of parameter space~\cite{pspace}. Assuming $\tilde\chi^0_1$ is 
the LSP, the 
production of $\tilde\chi^0_2$, either directly or indirectly from decays
of other supersymmetry particles, will result in events with two high
transverse energy ($E_T$) photons and large missing transverse 
momentum~($\rlap{\kern0.25em/}E_T$) and/or in events (\gmetnj) with one
high $E_T$ photon, multile jets and large \met. Moreover, the \ggmet\ events
are also expected in supersymmetric models with a light gravitino~($\tilde G$) 
being the LSP. In this case, $\tilde\chi^0_1$ (assumed to be the lightest
superpartner of a standard model particle) is unstable and decays into
a photon plus a gravitino ($\tilde\chi^0_1\rightarrow\gamma\tilde G$).

We present searches for new physics in the channels 
$p\bar{p}\rightarrow\gamma\gamma\rlap{\kern0.25em/}E_T+X$ and
$p\bar{p}\rightarrow\gamma\rlap{\kern0.25em/}E_T+\ge 2\ {\rm jets}+X$
at the Fermilab Tevatron. The $\gamma\rlap{\kern0.25em/}E_T$ events with 
fewer than two jets are not considered here, due to large backgrounds from
$W+$jets production. We interpret our results of the \ggmet\ analysis in 
terms of $\tilde e\tilde e$, $\tilde\nu\tilde\nu$, and 
$\tilde\chi^0_2\tilde\chi^0_2$ production and of the \gmetjj\ analysis in 
terms of squark ($\tilde q$) and gluino ($\tilde g$) production in 
supersymmetric models with a dominant 
$\tilde\chi^0_2\rightarrow\gamma\tilde\chi^0_1$ decay.
The interpretation of the results of the \ggmet\ analysis in models with 
a light $\tilde G$ can be found in Ref.~\cite{d0gg}.

The trigger used requires one electromagnetic (EM) 
cluster with transverse energy $E_T>15$~GeV, one jet with $E_T>10$~GeV, 
and $\rlap{\kern0.25em/}E_T$$>14$~GeV 
($\rlap{\kern0.25em/}E_T$$>10$~GeV for about 10\% of the data taken early 
in the Tevatron run). The jets in the trigger include non-leading EM clusters.
Photons are identified via a two-step process: the selection of isolated 
EM energy clusters and the rejection of electrons. The EM clusters are 
selected from calorimeter energy clusters by requiring
(i)   at least 95\% of the energy to be deposited in the EM section
      of the calorimeter,
(ii)  the transverse and longitudinal shower profiles to be consistent
      with those expected for an EM shower, and
(iii) the energy in an annular isolation cone with radius 0.2 to 0.4 around
      the cluster in $\eta-\phi$ space to be less than 10\% of the cluster
      energy, where $\eta$ and $\phi$ are the pseudorapidity and azimuthal
      angle.
Electrons are removed by rejecting EM clusters which have either a 
reconstructed track or a large number of tracking chamber hits in a road 
between the calorimeter cluster and the event vertex. 
$\rlap{\kern0.25em/}E_T$~is determined from the energy deposition in 
the calorimeter for $|\eta|<4.5$.

\subsection{Search for $\gamma\gamma\rlap{\kern0.25em/}E_T$ Events}
To be selected as $\gamma\gamma\rlap{\kern0.25em/}E_T$\ candidates, events
are first required to have two identified photons, one with 
$E_T^{\gamma_1}>20$~GeV and the other with $E_T^{\gamma_2}>12$~GeV,
each with pseudorapidity $|\eta^{\gamma}|<1.1$ or $1.5<|\eta^{\gamma}|<2.0$, 
the regions with good photon identification. We denote the 28 events passing 
these photon requirements as the $\gamma\gamma$ sample. We then require 
$\rlap{\kern0.25em/}E_T$$>25$~GeV with at least one reconstructed vertex 
in the event to ensure good measurement of $\rlap{\kern0.25em/}E_T$. 
No requirement on jets is made. Two events satisfy all requirements.
The data used in this analysis correspond to an integrated luminosity of
$106.3\pm 5.6$~pb$^{-1}$.

The principal backgrounds are multijet, direct photon, $W+\gamma$, 
$W+{\rm jets}$, $Z\rightarrow ee$, and $Z\rightarrow\tau\tau\rightarrow ee$ 
events from Standard Model processes with misidentified photons and/or 
mismeasured $\rlap{\kern0.25em/}E_T$.  The background due to 
$\rlap{\kern0.25em/}E_T$\ mismeasurement is estimated using events with 
two EM-like clusters which satisfy looser EM cluster requirements than 
those discussed above, and for which at least one of the two fails the 
EM shower profile consistency requirement (ii) above. In addition, 
these events must pass the photon kinematic requirements.  
By normalizing the number of events with $\rlap{\kern0.25em/}E_T$\ $<20$~GeV 
in the QCD sample to that in the $\gamma\gamma$ sample, we obtain a background 
of $2.1\pm 0.9$ events due to $\rlap{\kern0.25em/}E_T$\ mismeasurement
for $\rlap{\kern0.25em/}E_T$\ $>25$~GeV.

Other backgrounds are due to events with genuine $\rlap{\kern0.25em/}E_T$\
such as those from $W+$`$\gamma$' (where `$\gamma$' can be a real or a 
fake photon), $Z\rightarrow\tau\tau\rightarrow ee$, and
$t\bar{t}\rightarrow ee+{\rm jets}$ production. These events
would fake $\gamma\gamma\rlap{\kern0.25em/}E_T$\ events if the electrons
were misidentified as photons. We estimate their contribution using a
sample of $e$`$\gamma$' events passing the kinematic requirements, 
including that on $\rlap{\kern0.25em/}E_T$. Taking into account the 
probability ($0.0045\pm 0.0008$, determined from $Z\rightarrow e e$ data) 
that an electron is misidentified as a photon, we estimate a background 
of $0.2\pm 0.1$ events. Adding the two background contributions together 
yields $2.3\pm 0.9$ events. No excess of events is observed.

\begin{figure}[htb]
  \centerline{\epsfysize=3.0in\epsfbox{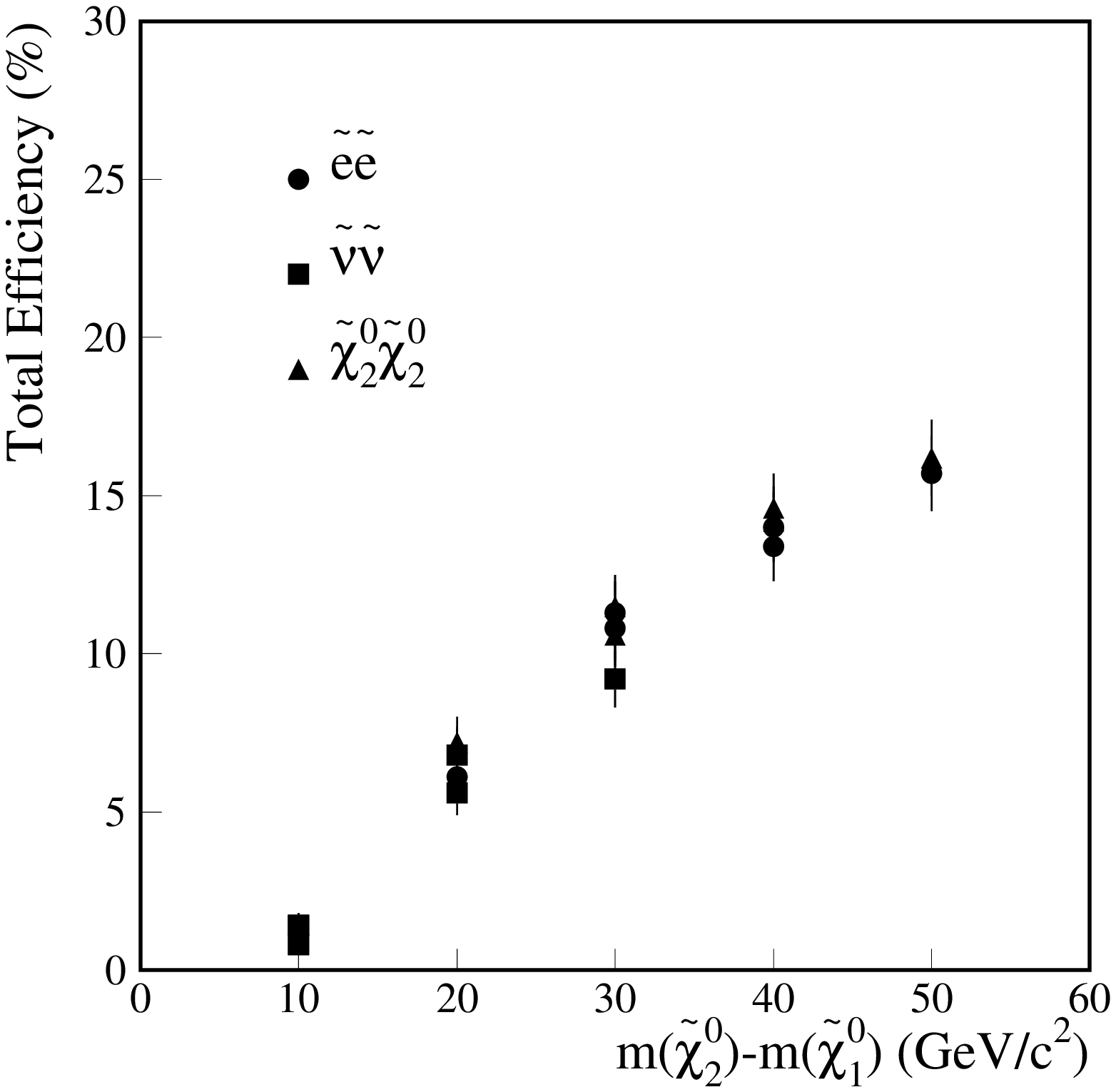}
              \epsfysize=3.0in\epsfbox{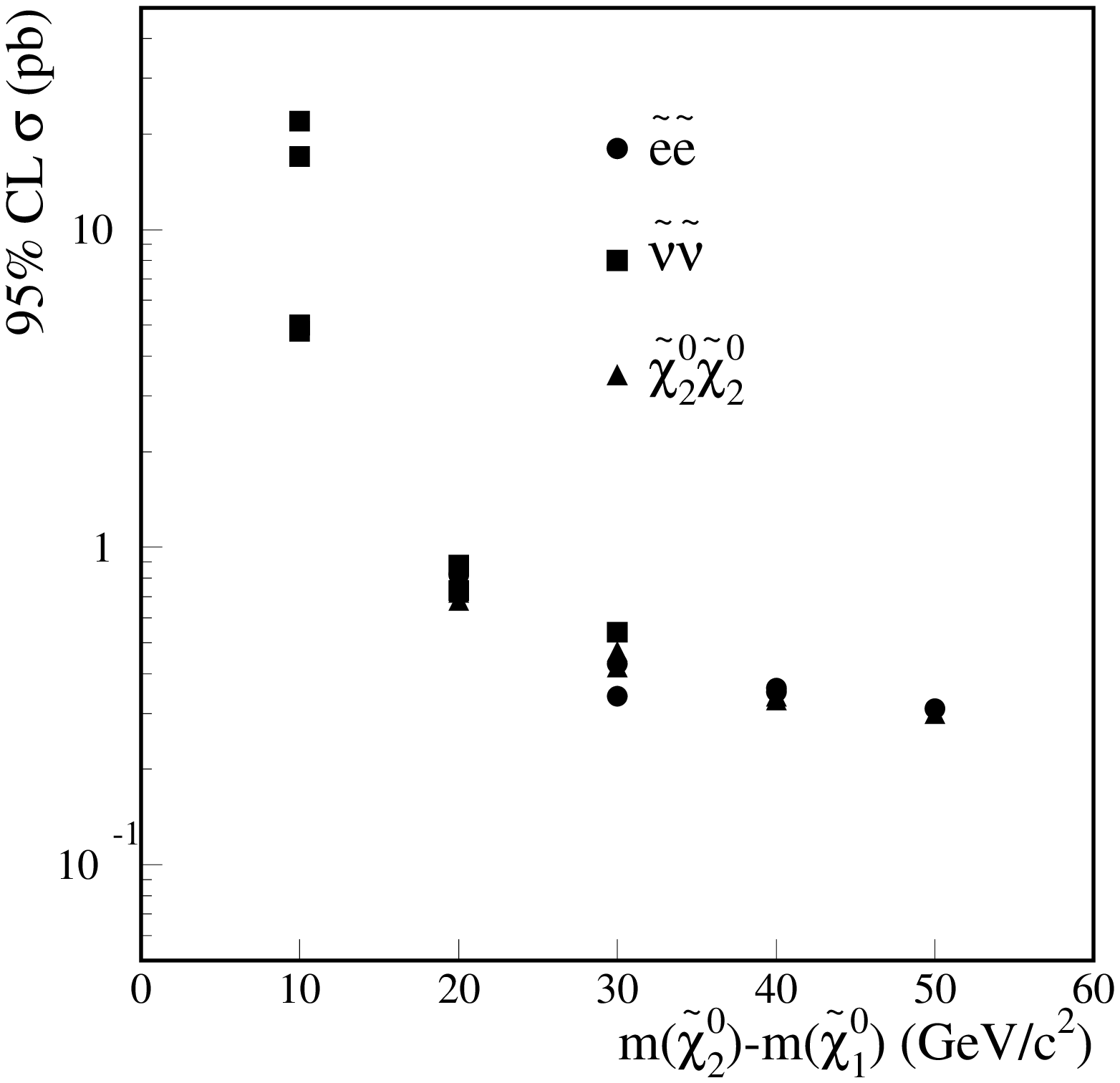}}
  \caption{(a) The total efficiency and (b) the 95\% C.L. 
               cross section limit as a function of 
               $m_{\tilde\chi^0_2}-m_{\tilde\chi^0_1}$ for 
               $p\bar{p}\rightarrow\tilde e\tilde e, \tilde\nu\tilde\nu, 
               \tilde\chi^0_2\tilde\chi^0_2$ processes. The error shown on 
               the efficiency is statistical only and the systematic error
               is estimated to be 6\%.}
  \label{fig:ggmet}
\end{figure}

The null results are used to set upper cross section limits on the pair 
production of scalar electron~($\tilde e$), scalar neutrino~($\tilde\nu$), and
$\tilde\chi^0_2$ with ${\rm Br}(\tilde e\rightarrow e\tilde\chi^0_2)=100\%$,
${\rm Br}(\tilde\nu\rightarrow\nu\tilde\chi^0_2)=100\%$, and 
${\rm Br}(\tilde\chi^0_2\rightarrow\gamma\tilde\chi^0_1)=100\%$.
$p\bar{p}\rightarrow\tilde e\tilde e, \tilde\nu\tilde\nu,
\tilde\chi^0_2\tilde\chi^0_2$ events are generated using the {\sc Isajet}
program~\cite{isajet} and are processed through the detector and 
trigger simulation and the reconstruction program. 
The $\tilde e$ mass is set to 100~GeV/c$^2$ while the $\tilde\nu$, 
$\tilde\chi^0_2$, and $\tilde\chi^0_1$ masses are varied
between 70--100~GeV/c$^2$, 50--90~GeV/c$^2$, and 30--80 GeV/c$^2$ respectively. 
The efficiency is found to be insensitive to the $\tilde e$ and $\tilde\nu$ 
masses and is largely determined by the mass difference 
$m_{\tilde\chi^0_2}-m_{\tilde\chi^0_1}$ as shown in Fig.~\ref{fig:ggmet}(a).
For a given value of $m_{\tilde\chi^0_2}-m_{\tilde\chi^0_1}$, the efficiencies
for the three processes are the same within the errors, independent of the 
values of $m_{\tilde e}$, $m_{\tilde\nu}$, and $m_{\tilde\chi^0_2}$.
The 95\% confidence level (C.L.) upper limits on the cross section
as a function of $m_{\tilde\chi^0_2}-m_{\tilde\chi^0_1}$ is shown in 
Fig.~\ref{fig:ggmet}(b). The limit is about 300~fb for large values of
$m_{\tilde\chi^0_2}-m_{\tilde\chi^0_1}$. These results improve over
those we published earlier~\cite{womer}, mainly due to the improved photon
identification. However, we note that
the expected cross sections for $p\bar{p}\rightarrow\tilde e\tilde e,
\tilde\nu\tilde\nu, \tilde\chi^0_2\tilde\chi^0_2$ processes are small, well
below the experimental upper limits. For example, the theoretical cross section 
for $p\bar{p}\rightarrow\tilde e_L\tilde e_L+\tilde e_R\tilde e_R$ production 
is about 30~fb for $m_{\tilde e}=100$~GeV/c$^2$.

\subsection{Search for $\gamma\rlap{\kern0.25em/}E_T+n$~Jets Events}

\begin{figure}[htb]
  \centerline{\epsfysize=3.0in\epsfbox{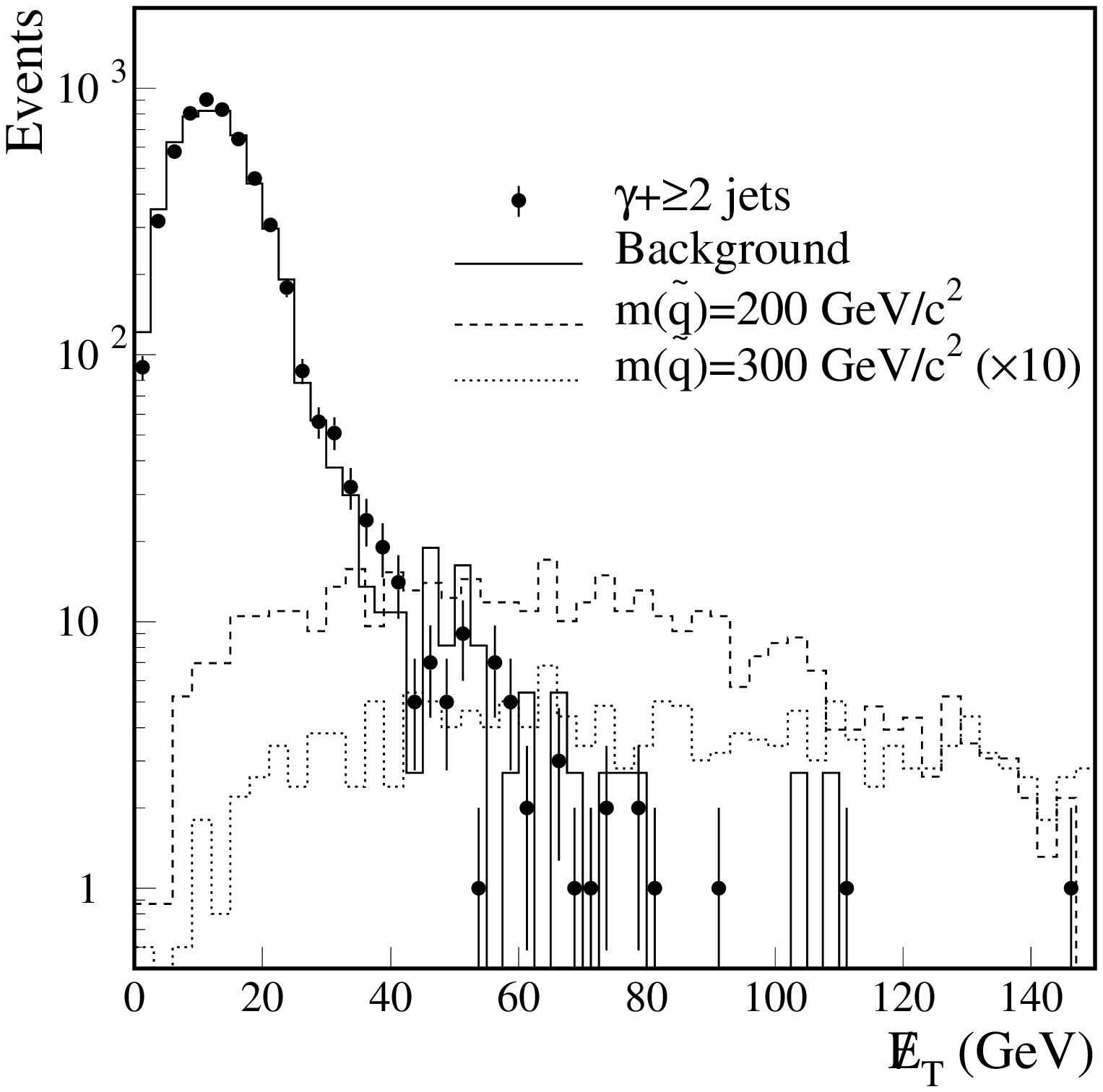}
              \epsfysize=3.0in\epsfbox{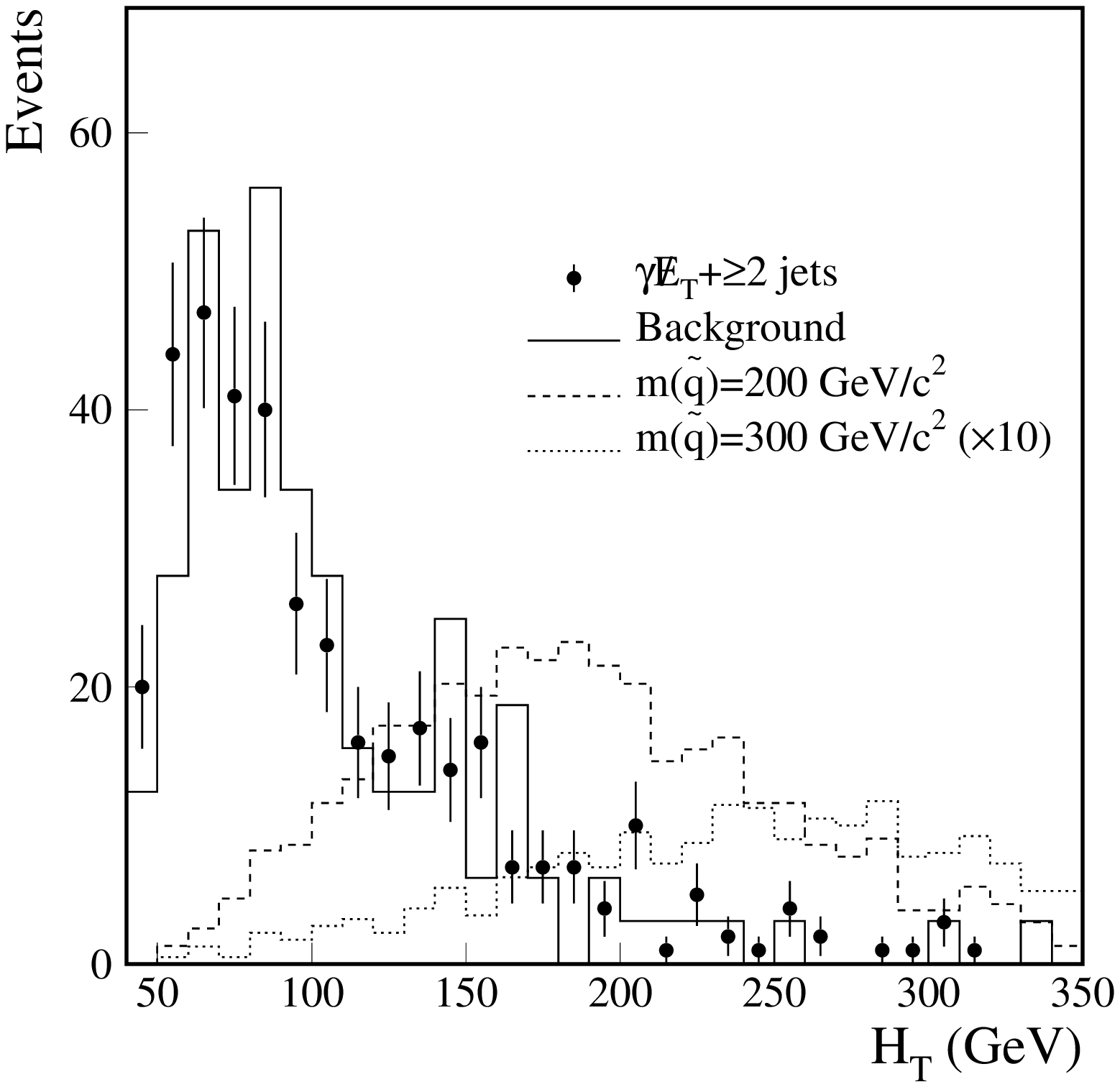}}
  \caption{(a) The \protect\met\ distribution (solid circles) of the events 
               with one photon and two or more jets.
               The expected \protect\met\ 
               distribution from the backgrounds is shown as a solid histogram.
               The number of events with \protect\met$<20$~GeV in the 
               background is normalized to that in the \protect\gjj\ sample.
           (b) The $H_T$ distributions of the \protect\gmetjj\ and background 
               events. 
           Also shown (dashed and dotted histograms) are the expected 
           distributions from supersymmetry 
           for two different values squark/gluino masses assuming an equal
           squark and gluino mass.}
  \label{fig:gmet_fig12}
\end{figure}

To be selected as \gmetjj\ candidates, events are first required to have at
least one identified photon with $E_T^\gamma>20$~GeV and pseudorapidity 
$|\eta^\gamma|<1.1$ or $1.5<|\eta^\gamma|<2.0$, and two or more jets with 
$E^j_T>20$~GeV and $|\eta^j|<2.0$. We denote events passing these requirements
as the \gjj\ sample. The \met\ distribution of the \gjj\ events is shown 
in Fig.~\ref{fig:gmet_fig12}(a). After requiring $\rlap{\kern0.25em/}E_T>25$~GeV, 
378 events remain in the sample. The data used in this analysis correspond to
an integrated luminosity of $99.4\pm 5.4$~pb$^{-1}$.

The principal backgrounds are multijet, direct photon, $W+{\rm jets}$, 
and $Z+{\rm jets}$ events from Standard Model processes with jets or electrons
misidentified as photons and/or mismeasured $\rlap{\kern0.25em/}E_T$.  
Following the same procedure as in the \ggmet\ analysis, the background due 
to $\rlap{\kern0.25em/}E_T$\ mismeasurement is estimated to be 
$370\pm 36$ events. $W+\ge 2\ {\rm jets}$ events with $W\rightarrow e\nu$ 
would fake \gmetjj\ events if the electrons were misidentified as photons. 
This contribution is estimated to be $4\pm 1$ using a sample of \emetjj\ 
events passing all the kinematic requirements with the electron satisfying 
those imposed on the photon. Another background due to 
$W(\rightarrow\ell\nu)+{\rm jets}$ and $Z(\rightarrow\nu\nu)+{\rm jets}$ 
production is found to be negligible. The total background ($374\pm36$)
agrees well with the number of observed events. The $H_T$ (defined as the 
scalar sum of the $E_T$ of all jets with $E_T>20$~GeV and $|\eta|<2.0$) 
distribution is shown in Fig.~\ref{fig:gmet_fig12}(b) for both \gmetjj\ 
and background samples. The background distribution reproduces the observed 
\gmetjj\ distribution well.

With the basic selection discussed above, the \gmetjj\ sample is dominated
by the backgrounds. To optimize the selection criteria for 
supersymmetry, we simulate squark and gluino production
(either in pair or in association with charginos/neutralinos) using the
{\sc Pythia} program~\cite{pythia}. The MSSM parameter values are set to 
$M_1=M_2=60.0$~GeV, $\tan\beta=2.0$, and $\mu=-40.0$~GeV.
This set of parameter values gives $m_{\tilde\chi^0_1}=33.5$~GeV/c$^2$,
$m_{\tilde\chi^0_2}=60.0$~GeV/c$^2$, and
${\rm Br}(\tilde\chi^0_2\rightarrow\gamma\tilde\chi^0_1)=100\%$.
Sleptons ($\tilde\ell$) and stop ($\tilde t$) are assumed to be heavy. 
Events with $\tilde\chi^0_2$ in the final state are selected and are run through
a detector simulation program, a trigger 
simulator, and the same trigger requirements, reconstruction, and analysis 
as the data. The \met\ and $H_T$ distributions for
$m_{\tilde q}(=m_{\tilde g})=200,300$~GeV/c$^2$ events after the 
basic selection are shown in Figs.~\ref{fig:gmet_fig12}.
The distributions expected from supersymmetry 
are considerably harder than those of the background events. To increase 
sensitivity to supersymmetry, we introduce an $H_T$ cut and maximize the 
$\epsilon/\sigma_b$ ratio by varying \met\ and $H_T$ requirements. Here 
$\epsilon$ is the efficiency for $m_{\tilde q}(=m_{\tilde g})=300$~GeV/c$^2$ 
MC events and $\sigma_b$ is the error on the estimated number of 
background events. The optimized cuts are found to be 
$\rlap{\kern0.25em/}E_T>45$~GeV and $H_T>220$~GeV. With these 
additional cuts, five \gmetjj\ events are observed while $8\pm 6$ events 
are expected from the background processes.

The efficiency for supersymmetry signal varies from a few percent for low mass 
$\tilde q/\tilde g$ to approximately 25\% for high mass $\tilde q/\tilde g$.
MC studies show that the efficiency varies by 4\% for 
different choices of $M_1$, $M_2$, $\tan\beta$, and $\mu$ which are 
consistent with ${\rm Br}(\tilde\chi^0_2\rightarrow\gamma\tilde\chi^0_1)=100\%$
and $m_{\tilde\chi^0_2}-m_{\tilde\chi^0_1}>20$~GeV/c$^2$ suggested in
Ref.~\cite{kane}. Experimentally, the mass requirement is needed to ensure 
that photons from $\tilde\chi^0_2$ decays are reasonably energetic and are 
detected with good efficiency. The variation is assigned as a systematic 
error in the efficiency. The total fractional systematic error on the 
efficiency is  9\%.

\begin{figure}[htbp]
  \centerline{\epsfysize=3.0in\epsfbox{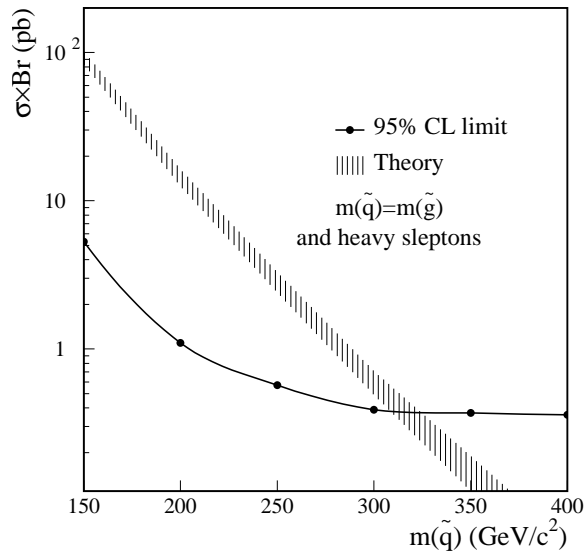}}
  \caption{The 95\% C.L. upper limit on the $\sigma\times{\rm Br}$ as a function
           of $m_{{\tilde q}/{\tilde g}}$ assuming an equal squark and gluino 
           mass. The hatched band represents the range of the theoretical 
           cross section for different sets of MSSM parameter values 
           consistent with the constraints 
           ${\rm Br}(\tilde\chi^0_2\rightarrow\gamma\tilde\chi^0_1)=100\%$ and
           $m_{\tilde\chi^0_2}-m_{\tilde\chi^0_1}>20$~GeV/c$^2$.}
  \label{fig:gmet_fig3}
%  \vspace*{-0.0in}
\end{figure}

With five events observed and $8\pm 6$ events expected from backgrounds, 
we observe no excess of events. We compute 95\% C.L. upper 
limits on $\sigma\times {\rm Br}=
\sigma(p\bar{p}\rightarrow\tilde q/\tilde g\rightarrow\tilde\chi^0_2+X)
\times {\rm Br}(\tilde\chi^0_2\rightarrow\gamma\tilde\chi^0_1)$
using a Bayesian
approach with a flat prior distribution for the signal cross 
section. The resulting upper 
limit as a function of squark/gluino mass is displayed in 
Fig.~\ref{fig:gmet_fig3}, for the case where 
$m_{\tilde q}=m_{\tilde g}$,  along with the theoretical cross sections, 
as calculated using the CTEQ3L parton distribution function 
(p.d.f.)~\cite{cteq}.
The hatched band represents the range of the theoretical cross sections
obtained by varying the supersymmetry parameter values with the constraints 
${\rm Br}(\tilde\chi^0_2\rightarrow\gamma\tilde\chi^0_1)=100\%$ and 
$m_{\tilde\chi^0_2}-m_{\tilde\chi^0_1}>20\ {\rm GeV/c}^2$. 
The intersection of our limit curve with the lower edge of the theory band 
is at $\sigma\times {\rm Br}=0.38$~pb, leading to a lower limit for equal mass
squarks and gluinos of 311~GeV/c$^2$.

The effect of light sleptons on the squark and gluino decays is studied 
by varying the slepton mass ($m_{\tilde\ell}$) in MC. The fraction 
of events containing at least one $\tilde\chi^0_2$ increases as the 
$m_{\tilde\ell}$ is decreased. When $m_{\tilde\ell}$ is varied from 
500~GeV/c$^2$ to 80~GeV/c$^2$, the fraction increases about 25\% 
for $m_{\tilde q}=m_{\tilde g}=300$~GeV/c$^2$ with equal gluino mass. 
Sleptons with mass below 80~GeV/c$^2$ have already been excluded~\cite{lep}.
The increased $\tilde\chi^0_2$ production increases the mass limit by 
approximately 10~GeV/c$^2$. 

A light stop ($\tilde t_1$) will also modify the squark and gluino decays 
and therefore affect the $\tilde\chi^0_2$ production. We investigate this 
effect by setting $m_{\tilde t_1}=80$~GeV/c$^2$ which approximately 
corresponds to the current $\tilde t_1$ lower mass limit~\cite{lep}. 
A 15\% reduction in 
$\tilde\chi^0_2$ production cross section is observed. This reduction 
lowers the limit for equal mass squarks and gluinos by about 6~GeV/c$^2$.

Following the procedure above, we obtain a low mass limit for gluinos (squarks) 
to be 233~GeV/c$^2$ (219~GeV/c$^2$) when squarks (gluinos) are heavy. 
Again, these limits vary approximately 10~GeV/c$^2$ if $\tilde t_1$ and/or 
sleptons are light.

\section{Search for Dirac Monopoles}
One of the  open questions  of particle  physics is the  existence of Dirac
monopoles~\cite{Dirac,Schwinger},    hypothetical carriers  of the magnetic
charge proposed by P.M.~Dirac to symmetrize Maxwell equations and explain
the quantization of  electric charge. If  such magnetic monopoles exist,
then the  elementary magnetic  and electric  charges ($g$  and $e$) must be
quantized according to the following formula:
\vspace*{-0.05in}
\begin{equation}
    g = \frac{2\pi n}{e},\;\; n = \pm 1, \pm 2, ...,
\vspace*{-0.05in}
\end{equation}
where $n$ is an unknown  integer. 

\begin{figure}[htb]
  \centerline{\epsfysize=2.0in\epsfbox{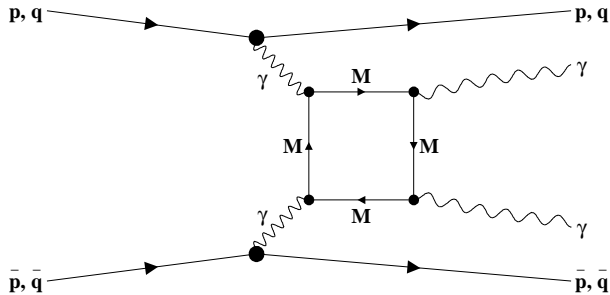}}
  \caption{Feynman diagram for $\gamma\gamma$ production via a virtual monopole
           loop.}
  \label{fig:mpole_diagram}
  \vspace*{-0.0in}
\end{figure}

Dirac   monopoles are  expected  to  couple to  photons  with a
coupling constant $\alpha_g =  g^2/4\pi \approx 34\; n^2$ which is at least
three orders of magnitude larger  than the corresponding photon coupling to
the electric charge  ($\alpha_e = e^2/4\pi  \approx 1/137$). Therefore such
monopoles could give rise to photon-photon rescattering via the box diagram
shown in Fig.~\ref{fig:mpole_diagram}~\cite{Ginzburg1,DeRujula}. 
The contribution
of this diagram  for pointlike  monopoles to diphoton  production at hadron
colliders  was   recently     calculated~\cite{Ginzburg2}  and  shown to be
significant  even  for  monopole  masses  comparable to  the  collider beam
energy.

Since the  virtuality  ($Q^2$) of most  incoming photons  in the process of
Fig.~\ref{fig:mpole_diagram} is small~\cite{Ginzburg-private}, the interacting
partons  scatter at very  small  angles and  therefore escape  the detector
through the beam pipe. 
Thus, a  signature for monopoles at hadron colliders
is the   production of a  pair of  isolated  photons  with high  transverse
energies.  This process  gives a  unique  opportunity to find  evidence for
Dirac  monopoles  or to  set  limits on the  monopole  mass.
Previous monopole searches can be found in Ref.~\cite{L3,PDG}.

Despite  numerous  studies,  QED with  pointlike  monopoles is  still not a
complete theory. For example, it is  not clear whether such a theory can be
constructed  to be   renormalizable to  all   orders~\cite{DeRujula}. Also,
arguments  exist (see,   e.g.~\cite{Goldhaber})  that Dirac  monopoles must
occupy a spatial  volume of radius  $R \sim {\cal  O}(g^2/M)$, where $M$ is
the  monopole mass, to  accommodate  the self  energy implied  by the large
coupling. In such a theory, it is  possible that hard  interactions of a  
monopole with photons would be weakened substantially by the effects of a 
monopole form factor.

The data used in this analysis represent an integrated luminosity of 
$69.5\pm 3.7$~pb$^{-1}$ using a trigger which required the presence of
an electromagnetic object with transverse  energy $E_T$ above 40~GeV.
This trigger did  not require the  presence of an  inelastic collision, and
therefore can be used to select low  $Q^2$ events typical of the process in
Fig.~\ref{fig:mpole_diagram}.

The following offline selection criteria are: (i) at least two photons with
$E_T >  40$~GeV and   pseudorapidity  $|\eta^\gamma|  < 1.1$;  (ii) missing
transverse  energy in  the event  \met$<25$~GeV; and (iii)  no jets with
$E_T^j > 15$~GeV  and  $|\eta^j|<2.5$. The jet veto  requirement is used to
select the  low $Q^2$  process in   Fig.~\ref{fig:mpole_diagram}.  
The trigger is $>98$\% efficient for this off-line selection.

Photons are selected from the identified EM clusters by requiring no tracks
pointing toward the cluster from any of the event vertices.
The overall efficiency for photon  identification is $(73.0 \pm 1.2)\%$ per
photon. The total efficiency for diphoton is $(52.8 \pm 1.4)\%$. This 
includes the   efficiency of  the  \met\ veto  $(99.0  \pm  0.5)$\% as well 
as the identification efficiency for a pair of photons. The  above  selection 
criteria  define  our base  sample which  contains 90 candidate events.

The main backgrounds to  photon scattering  through a monopole loop are due
to: (i) diagrams similar to  Fig.~\ref{fig:mpole_diagram} with other particles
in the  loop;  (ii)  QCD  production  of  dijets  ($jj$)  and  direct  photons
($j\gamma$) (with jets misidentified as photons due to fragmentation into a
leading $\pi^0$ or $\eta$ decaying  into a pair of spatially close photons,
reconstructed as one EM cluster), or direct diphotons ($\gamma\gamma$); and
(iii)  Drell-Yan  dielectron  production  with  electrons  misidentified as
photons due to tracking inefficiency.

Background (i) is dominated by a  virtual $W$-loop and has been shown to be
negligible~\cite{W-loop}.    The  other two  background   contributions are
estimated  from  the  data. The  QCD  background  is  determined  using the
$j\gamma$ event sample collected with a single photon trigger, with the jet
passing the same  fiducial and  kinematic cuts as the  photon. By applying 
a jet-faking probability, we find  the QCD  background to be  $25 \pm 8$  
events.  Direct photon and diphoton  backgrounds are  also included in  
this estimate.  

The Drell-Yan  background is calculated from a  sample of dielectron events
passing the same fiducial and kinematic cuts as the signal sample. Multijet
contamination of this sample is  negligible. The  probability  for a  
dielectron pair  to be  misidentified as a
diphoton  pair is found  to be $(11  \pm 1)\%$ by  comparing  the number of
events in the $Z$ peak in the $ee$ and $\gamma\gamma$ samples passing loose
kinematic cuts. The  Drell-Yan background in the  base sample is $63 \pm 7$
events. The  overall  background in the base  sample is $88  \pm 11$~(syst)
events, in good agreement with the 90 observed candidates. 

\begin{figure}[htb]
  \centerline{\epsfysize=2.0in\epsfbox{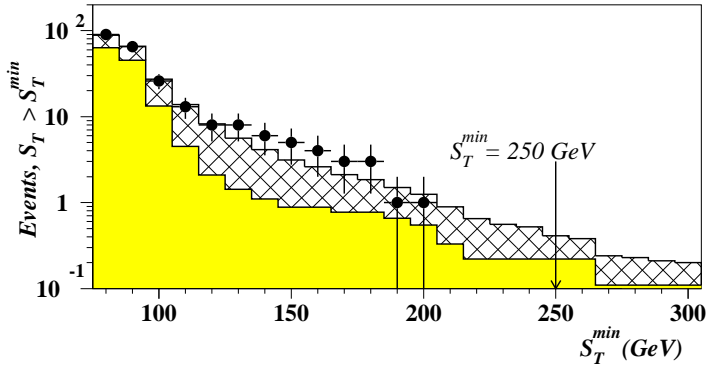}}
  \caption{Data and  expected  background  as a  function of  $S_T^{\rm min}$
           cut. Points are data, the upper hatched region corresponds to the 
           QCD background, and the lower shaded region shows the Drell-Yan 
           background. The $\approx 15$\% systematic error on the background 
           is not shown.} 
  \label{fig:mpole_bck}
  \vspace*{-0.0in}
\end{figure}

To   optimize  the   sensitivity  of  this  search  to  the   monopole loop
contribution we apply a cut on the scalar sum of the transverse energies of
all the photons in the event: $S_T \equiv \sum_i E_T^{\gamma_i}$.
We vary  the  $S_T$ cut  threshold   ($S_T^{\rm min}$)  in 10  GeV steps to
achieve   an  expected background of  0.4 events.  Such an
optimization is based on the fact that for this expected background one has
a 67\%  probability of  observing no  candidate  events in  the data in the
absence of a signal. In such a case~\cite{PDG}, the limits on the signal do
not  depend on  the  exact   background  value or  its   uncertainties. The
agreement   between  the   observed  number of   events and  the  predicted
background    as a   function  of    $S_T^{\rm  min}$  is    illustrated in
Fig.~\ref{fig:mpole_bck}. 
The $S_T^{\rm min} = 250$~GeV cut  corresponds to a background of $0.41 \pm
0.11$   events. We set an upper limit for the production cross section of two or
more photons with $\sum E_T^\gamma > 250$~GeV and $|\eta^\gamma| < 1.1$: 
\begin{equation}	
    \sigma(p\bar p \to \ge 2\gamma)|_{S_T >  
    250~\mbox{\tiny GeV},  |\eta^\gamma| < 1.1} < 83~\mbox{fb} 
\end{equation} 
at the 95\% C.L.  

Since the data  are consistent with  the background  hypothesis, we can set
limits on the  production of  pointlike  Dirac monopoles.  We calculate the
acceptance for the  monopole signal using a fast  MC program that generates
diphoton   events  from  a  monopole  loop   according  to  the  calculated
differential   cross  section     $d^3\sigma/dE_T^\gamma   d\eta^{\gamma_1}
d\eta^{\gamma_2}$~\cite{Ginzburg2}  with a subsequent parametric simulation
of the  D\O\  detector. 
The overall acceptance for the monopole signal is found to be $(51 \pm 1)\%$.
The acceptance does not depend on
the monopole mass  for masses above  the typical photon  energy ($\sim 300$
GeV)~\cite{Ginzburg-private}.

The total cross section for heavy monopole production at the Tevatron is
given by~\cite{Ginzburg2}:
\vspace*{-0.1in}
\begin{equation}
    \sigma(p\bar p \to \gamma\gamma + X) = 57\,P\; \left(\frac{n}{M~[{\rm
    TeV}]}\right)^8~{\rm fb,}
\end{equation}
where $P$ is  a spin  dependent   factor~\cite{W-loop,P-factor}: P = 0.085,
1.39,  and 159  for  monopole  spin of  0, 1/2,  and 1,   respectively. The
estimated error on this cross section due to choice of p.d.f.
 and to higher
order QED effects is 30\%~\cite{Ginzburg-private}. Additional uncertainties
are  associated with  the  $\gamma\gamma  \to  \gamma\gamma$  subprocess in
Fig.~\ref{fig:mpole_diagram}   and with  unitarity  considerations.  The coupling
constant $\alpha_g$ is replaced with an effective coupling~\cite{Ginzburg2}
obtained by  multiplying  $\alpha_g$  by a factor   $(E^\gamma/M)^2$, where
$E^\gamma$ is the  photon energy,  typically 300~GeV at  the Tevatron. Both
unitarity  and  perturbation  theory  assumptions  are  satisfied when this
factor is $\ll$~1~\cite{Ginzburg1,Ginzburg2}.

\begin{figure}[htb]
  \centerline{\epsfysize=3.0in\epsfbox{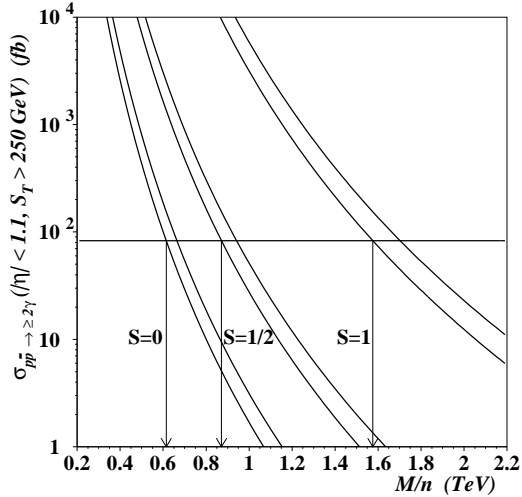}}
  \caption{The curved bands show the low and upper bounds on theoretical cross
           sections~\protect\cite{Ginzburg2} for monopole spin, $S = 0, 1/2,$ 
           and 1. The horizontal line shows the 95\% C.L. experimental upper 
           limit on the cross section. The arrows 
           indicate the lower 95\% C.L. limits on the monopole mass at each spin
           value.}
  \label{fig:mpole_limits}
%  \vspace*{-0.5in}
\end{figure}

Comparing the  lower bound of the  theoretical cross  section corrected for
acceptance  with  the cross  section  limit   set by this
measurement, we  obtain the  following lower limits on  the pointlike Dirac
monopole mass (see Fig.~\ref{fig:mpole_limits}): 
\vspace*{-0.1in}
$$
M/n > \left\{
    \begin{array}{rl}
     610~\mbox{GeV} & \mbox{for~} S = 0 \\
     870~\mbox{GeV} & \mbox{for~} S =  1/2  \\ 
    1580~\mbox{GeV} & \mbox{for~} S =  1
    \end{array}\right..  
$$

We note that the effective coupling exceeds 1 and unitarity is violated
close to the  experimental  bound. For  values $E^\gamma/M  > 1$, the cross
section will grow more slowly,  approaching the usual $1/M^2$ behavior of a
QED  process~\cite{Ginzburg-private}  which satisfies  unitarity. Also, for
lower monopole  masses the effective  parameter of the  perturbation theory
used in the  calculations~\cite{Ginzburg2} becomes too large, and therefore
one would expect a non-negligible contribution of the higher order diagrams
with four,  six, etc.  photons in  the final  state. The  latter effect is,
however, largely  compensated by our analysis cut  on the sum of the photon
transverse  energies; if  part of  the signal  cross section  is due to the
higher order diagrams, the above limits are unaffected.

\section{Summary}

In summary, we have performed searches for supersymmetry by searching for
\ggmet\ and \gmetjj\ events and for heavy pointlike Dirac monopoles by 
searching for pairs of  photons with high transverse energies. Our data agree 
with the expected  background production. Within the framework of the MSSM 
with the choices of the parameter values consistent with 
${\rm Br}(\tilde\chi^0_2\rightarrow\gamma\tilde\chi^0_1)=100\%$ and 
$m_{\tilde\chi^0_2}-m_{\tilde\chi^0_1}>20$~GeV/c$^2$, we obtain a 95\% C.L. 
lower mass limit of 311 GeV/c$^2$ for equal mass squarks and gluinos.
Using theoretical calculations~\cite{Ginzburg2}  we  set 95\% C.L. lower 
limits on the Dirac monopole  mass for  minimum  magnetic  charge  $(n=1)$ 
in the  range 610 to 1580~GeV, depending on the monopole spin.


\begin{thebibliography}{99}

  \bibitem{dzero}
     D\O\ Collaboration, S. Abachi {\it et al.}, Nucl. Instrum. Methods A
                         {\bf 338}, 185 (1994).
  \bibitem{susy}
    For a review, see for example: 
    H.E.~Haber and G.~Kane, Phys. Rept. {\bf 117}, 75 (1985).

  \bibitem{gmsb}
     S.~Dimopoulos, S.~Thomas, and J.D.~Wells,
                   Phys. Rev. D {\bf 54}, 3283 (1996);
     S.~Dimopoulos {\it et al.}, Phys. Rev. Lett. {\bf 76}, 3494 (1996);
     K.S.~Babu, C.~Kolda, and F.~Wilczek,
                   Phys. Rev. Lett. {\bf 77}, 3070 (1996);
     J.L.~Lopez, D.V.~Nanopoulos, and A.~Zichichi,
                  Phys. Rev. Lett. {\bf 77}, 5168 (1996);
     S.~Ambrosanio {\it et al.}, Phys. Rev. D {\bf 54}, 5395 (1996);
     H.~Baer {\it et al.}, Phys. Rev. D {\bf 55}, 4463 (1997);
     J.~Ellis, J.L.~Lopez, and D.V.~Nanopoulos, Phys. Lett. B 
     {\bf 394}, 354 (1997).

  \bibitem{kane}
    S. Ambrosanio {\it et al.}, Phys. Rev. Lett. {\bf 76}, 3498 (1996) and
                                Phys. Rev. D {\bf 55}, 1372 (1997).

  \bibitem{cdf}
     CDF Collaboration, F. Abe {\it et al.}, {\tt hep-ex/9801019}, 
        submitted to Phys. Rev. Lett.

  \bibitem{pspace}
    S. Ambrosanio and B. Mele, Phys. Rev. D {\bf 55}, 1399 (1997), 
             Erratum-ibid. D {\bf 56}, 3157 (1997).

  \bibitem{d0gg}
    D\O\ Collaboration, B. Abbott {\it et al.}, Phys. Rev. Lett. 
                         {\bf 80}, 442 (1998).

  \bibitem{isajet}
     F.E. Paige and S.D. Protopopescu, BNL report No. BNL38034 (1986)
     (unpublished).

  \bibitem{womer}
     D\O\ Collaboration, S. Abachi {\it et al.}, Phys. Rev. Lett. {\bf 78},
                         2070 (1997).

  \bibitem{pythia}
     H.U. Bengtsson and T. Sj\"ostrand, Comp. Phys. Comm. {\bf 46}, 43 (1987);
     T. Sj\"ostrand, Comp. Phys. Comm. {\bf 82}, 74 (1994);
     S. Mrenna, Comp. Phys. Comm. {\bf 101}, 232 (1997).

  \bibitem{cteq}
     H. L. Lai {\it et al.}, Phys. Rev. {\bf D51}, 4763 (1995).

  \bibitem{lep}
     D\O\ Collaboration, S. Abachi {\it et al.}, Phys. Rev. Lett. {\bf 76}, 2222 (1996);
     A.~Kounine, ``{\em Higgs and Supersymmetry Searches at LEP}'', 
                 see this proceeding.

  \bibitem{Dirac}
     P.A.M.~Dirac, Proc. R. Soc. London, Ser. {\bf A} 133, 60 (1931).

  \bibitem{Schwinger}
     J.~Schwinger, Phys. Rev. {\bf 151}, 1055 (1966).

  \bibitem{Ginzburg1}
     I.F.~Ginzburg, S.L.~Panfil, Sov. J. Nucl. Phys. {\bf 36}, 850 (1982).

  \bibitem{DeRujula}
     A.~De R\'ujula, Nucl. Phys. {\bf B435}, 257 (1995).

  \bibitem{Ginzburg2}
     I.F.~Ginzburg and A.Schiller, {\tt hep-ph/9802310}, 
          To appear in Phys. Rev. D.
  
  \bibitem{Ginzburg-private}
     I.F.~Ginzburg, private communication.

  \bibitem{L3}
     L3 Collaboration, M.~Acciarri {\it et al}, Phys. Lett. B {\bf 345}, 609 (1995).

  \bibitem{PDG}
     PDG Review of Particle Physics, Phys. Rev. D {\bf 54}, 166, 685-687 (1996).

  \bibitem{Goldhaber}
     A.S.~Goldhaber, in Proceedings of the CRM--FIELDS--CAP Workshop 
     ``Solitons'' at Queen's University, Kingston, Ontario, July 1997
     (Springer, New York 1998).

%  \bibitem{Zvvg}
%     D\O\  Collaboration, S.~Abachi  {\it et al}, Phys. Rev. Lett. {\bf 78}, 3640
%      (1997); {\it ibid.} Phys. Rev. D {\bf 56}, 6742 (1997).
  
  \bibitem{W-loop}
     G.~Jikia and A.~Tkabaladze,    Phys. Lett.  B {\bf 323},  453  (1994).

%  \bibitem{GRV}
%     M.~Gl\"uck, E.~Reya and A.~Vogt, Z. Phys. {\bf C67}, 433 (1995).

  \bibitem{P-factor}
     W.~Heisenberg and H.~Euler, Z. Phys.  {\bf 38}, 714 (1936); V.~Constantini,
     B. De Tollis, and G.~Pistoni, Nuovo Cim. {\bf 2A}, 733 (1971);   M.~Baillagreon,
     G.~Belanger,   and    F.~Boudjema,  Phys.  Rev. D  {\bf  51},  4712  (1995);
     M.~Baillagreon,  F.~Boudjema, E.~Chopin, and  V.~Lafage, Z. Phys. C {\bf 67},
     431 (1996).

\end{thebibliography}
\end{document}